\documentclass[twoside,11pt]{article}
\usepackage{graphicx}
\usepackage{bm}
\usepackage{booktabs}
\usepackage[square, numbers]{natbib}
\usepackage[margin=1in]{geometry}
\usepackage[usenames]{color}
\usepackage{float}
\usepackage{amsfonts, amssymb, amsmath, amsthm, multicol}
\usepackage{url}
\usepackage[colorlinks=true, pdfstartview=FitV, linkcolor=blue,
citecolor=blue, urlcolor=blue]{hyperref}
\usepackage[usenames]{color}
\definecolor{Red}{rgb}{1,0.05,0}
\definecolor{Grn}{rgb}{0.1,0.7,0.1}
\definecolor{Blu}{rgb}{0.1,0.1,0.6}
\definecolor{Org}{rgb}{1,0.45,0}
\definecolor{Vio}{rgb}{0.6578,0,0.9478}
\definecolor{Mag}{rgb}{1,0.2,0.3}
\usepackage{authblk}
 \usepackage{booktabs}
 \usepackage{lineno}
 \usepackage[section]{placeins}
\newtheoremstyle{mythmstyle}%
    {\topsep}
    {\topsep}
    {\itshape\setlength{\parskip}{0pt}\setlength{\parindent}{1.5em}}
    {\parindent}
    {\bfseries}
    {:}
    {0.5em}
    {}
\usepackage{upgreek}

\usepackage{xcolor}
\usepackage{soul}
\usepackage{graphicx}
\usepackage{caption}
\usepackage{float}
\usepackage{dcolumn}
\usepackage{bm}
\usepackage{upgreek}
\usepackage{lineno}
\usepackage[ruled, noline, linesnumbered, commentsnumbered, longend]{algorithm2e}
\usepackage{subcaption}
\usepackage{pgfplots}
\usepackage{xcolor}
\usepackage{tikz}
\usepgfplotslibrary{groupplots}
\pgfplotsset{compat=1.18}
\usetikzlibrary{calc}
\pgfplotsset{compat=newest}
\usepackage{pgfplots}
\usepackage{amsmath}
\usepackage[mathscr]{euscript}
\usepackage{booktabs}   
\usepackage{multirow}   
\usepackage{adjustbox}
\usepackage{subcaption}
\usepackage{hyperref}
\newcommand{\ulX}{\underline{\mathbf{X}}}
\newcommand{\ulG}{\underline{\mathbf{G}}}

\pgfplotsset{
  uniform plot font/.style={
    tick label style={font=#1},
    label style={font=#1},
    legend style={
      font=#1,
      nodes={font=#1}
    }
  }
}

\newcommand{\ultheta}{{\underline{\bm{\Uptheta}}}}

\newcommand{\ulhattheta}{\mathbf{\underline{\bm{\widehat{\Uptheta}}}}}
\usepackage{colortbl}

\title{Low-Rank Tensor Manifolds for Turbulent Field Reconstruction from Sparse Data}

\author[1]{Nazmus Sakib}
\author[2]{Alexander G. Mychkovsky}
\author[1]{Hessam Babaee\footnote{Corresponding author: h.babaee@pitt.edu}}

\affil[1]{Department of Mechanical Engineering and Materials Science, University of Pittsburgh, Pittsburgh, PA 15261, USA}
\affil[2]{Naval Nuclear Laboratory,
West Mifflin-PA and Schenectady-NY, USA}

\date{}

\begin{document}
\maketitle \vspace{1cm}

\begin{abstract}
Reconstructing turbulent flow fields from sparse and irregular measurements remains challenging because the underlying inverse problem is severely underdetermined. Here we present MANTRA, a manifold-based tensor framework that represents each velocity component in a continuous tensor-product space while constraining its coefficient tensor to a low-rank Tucker manifold. This structure provides a compact, data-adaptive representation without requiring gridded observations or governing equations. Riemannian optimization and adaptive selection of rank and basis complexity enable the model to exploit coordinate-wise spatiotemporal correlations. In simulations of forced isotropic turbulence and flow over a cylinder, MANTRA recovers multiscale structures at sampling densities orders of magnitude below the underlying grids and remains accurate under Gaussian noise up to 20\% of the peak velocity. It also recovers leading coordinate-wise correlation structures from sparse observations. Applied to experimental particle-tracking velocimetry of an oscillating impinging jet, MANTRA reveals time-evolving structures that are not apparent from the scattered measurements alone.
\end{abstract}

\section{Introduction}

Turbulent flows are among the most challenging systems to measure, model, and reconstruct. Despite advances in sensing and simulation, measurements remain sparse relative to the enormous number of degrees of freedom needed to resolve multiscale turbulent dynamics. The challenge is particularly acute when measurements are irregular, noisy, spatially sparse, or temporally incomplete, rendering flow reconstruction a severely underdetermined inverse problem~\citep{bui2004aerodynamic,carrassi2018data}. Lagrangian particle tracking velocimetry (LPTV) exemplifies this challenge. As one of the leading techniques for volumetric flow measurements, it yields sparse, irregularly distributed particle trajectories and velocities in complex three-dimensional unsteady flows~\citep{maas1993particle,virant19973d,ouellette2006quantitative,Dabiri_ptv,schanz2016shake}. Recovering continuous, or at least finely resolved, spatiotemporal velocity fields from these sparse observations is essential for computing velocity gradients, vorticity, pressure, coherent structures, and transport statistics that underpin fluid dynamics~\citep{schanz2016shake,schneiders2016dense,schroder20233d}.

The central challenge is to infer the underlying spatiotemporal velocity field from limited observations. Representing multiscale turbulent dynamics requires highly expressive models whose degrees of freedom typically far exceed the number of available measurements, rendering the reconstruction problem severely underdetermined, ill-conditioned, and non-unique. From a statistical learning perspective, successful reconstruction therefore requires an inductive bias that restricts the space of admissible solutions. Such inductive biases can arise from regularization, prior models, architectural constraints, or physical structure, each restricting the hypothesis space while promoting statistically stable and physically plausible reconstructions. The effectiveness of any reconstruction method therefore depends critically on the choice of inductive bias.

Existing approaches address the ill-posedness of flow reconstruction by introducing inductive biases through regularization. Examples include ridge ($\ell_2$) regularization, which favors small-norm, low-energy solutions~\citep{JAL19}; sparsity-promoting ($\ell_1$) regularization, which exploits sparse representations~\citep{C06,CMB19}; and physics-informed constraints that enforce governing equations and conservation laws within learning and data-assimilation frameworks~\citep{RPK19,CWF21,jeon2022fine}. Although these approaches differ in implementation, they all restrict the set of admissible solutions to reduce the ill-posedness inherent in severely underdetermined reconstruction problems.

Another successful strategy exploits coherent structures by representing
the flow in a low-dimensional modal subspace, reducing reconstruction to
the estimation of a small set of modal coefficients. This paradigm
underlies proper orthogonal decomposition (POD), gappy POD, and related
sparse-sensing approaches~\citep{everson1995karhunen,venturi2004gappy,willcox2006unsteady,CMB19,RCV12}, which represent the flow using global
modes learned from representative full-field data. Once these modes are
available, sparse measurements can be used to estimate the modal
coefficients and reconstruct the flow. This offline--online strategy has
also been extended to tensor-based representations that preserve the
multidimensional structure of the field and reconstruct it from optimally
placed sensors~\citep{FS23}. The problem becomes considerably
more challenging when the low-dimensional representation itself is
unknown and must be inferred directly from sparse measurements. In
particle-tracking velocimetry and related Lagrangian sensing modalities,
measurements consist of scattered particle locations that vary from frame
to frame, making the direct construction of correlation operators—and
consequently the extraction of modal or tensor subspaces—substantially
more difficult.

In this work, the ambient representation is constrained to a low-rank Tucker manifold, providing a geometric prior for the reconstruction problem. Restricting the solution to this low-dimensional manifold regularizes the otherwise underdetermined inverse problem, enabling accurate estimation even when the number of measurements is substantially smaller than the number of ambient parameters. Unlike POD-based approaches, which primarily exploit temporal correlations through global modes, the Tucker manifold captures coordinate-wise low-dimensional structure across both space and time. Consequently, accurate reconstruction does not require long time histories or strongly correlated snapshots and remains effective over short observation windows, including single-snapshot settings. Because the representation is learned directly from scattered measurements, the method naturally accommodates sparse, irregular observations without requiring interpolation onto a common Eulerian grid. The manifold formulation further enables a geometric analysis of the reconstruction problem together with efficient Riemannian optimization algorithms that operate directly on the low-rank solution manifold.

More broadly, this work is motivated by the growing success of low-rank tensor methods in scientific computing and machine learning. Originally developed in many-body quantum physics and quantum chemistry through tensor-network representations of high-dimensional wavefunctions~\citep{W92,Beck:2000aa,O14}, these methods have driven major advances in the mathematical foundations, numerical analysis, and algorithms of low-rank approximation~\citep{KB09,KL07,TT11,GKT13,B23,DPNFB23,TKZ26}. More recently, low-rank matrix and tensor-network methods have found widespread application in high-dimensional and stochastic PDEs~\citep{K12,DKS21,GBTT24}, turbulent flow simulation~\citep{RNB21,GLD22,petal24,AB24}, and deep learning~\citep{NPOV15,SZK22}. Here, we extend these ideas to a different setting by using low-rank tensor geometry as a data-driven inductive bias for reconstructing turbulent flow fields from sparse measurements, without requiring access to the governing equations.
\section{Methods}\label{Sect.Method}
\subsection{Notation and Basic Definitions}\label{Sect.Noation}

We use regular letters (non-bold) for scalars (e.g., $x$), bold lowercase
letters for vectors (e.g., $\mathbf{x}$), bold uppercase letters for matrices
(e.g., $\mathbf{X}$), and underlined bold uppercase letters for tensors
(e.g., $\ulX$). The identity matrix of size $n$ is denoted by $\mathbf{I}_n$. 

A $d$-way tensor
$ \ulX \in \mathbb{R}^{p_1\times\cdots\times p_d}$
belongs to an \emph{ambient} Euclidean tensor space with dimension $P=\prod_{k=1}^{d} p_k.$
The mode-$k$ unfolding of $\ulX$, denoted by $\mathbf{X}_{(k)}
\in
\mathbb{R}^{p_k\times\prod_{j\neq k}p_j},$
is obtained by arranging the $k$th mode as rows and collapsing all remaining
modes into columns. The multilinear rank of $\ulX$ is the tuple
\[
\operatorname{rank}_{\mathrm{ML}}(\ulX)
=
\left(
\operatorname{rank}(\mathbf{X}_{(1)}),
\ldots,
\operatorname{rank}(\mathbf{X}_{(d)})
\right).
\]

For a tensor with multilinear rank
$\mathbf{r}=(r_1,\ldots,r_d)$, a Tucker representation is given by
\begin{equation}\label{eq:Tucker}
\ulX
=
\ulG
\times_1 \mathbf{U}_1
\times_2 \mathbf{U}_2
\cdots
\times_d \mathbf{U}_d,
\end{equation}
where
$\ulG\in\mathbb{R}^{r_1\times\cdots\times r_d}$
is the core tensor and $\mathbf{U}_k\in\mathbb{R}^{p_k\times r_k},
\  k=1,\ldots,d,$
are factor matrices. Without loss of generality, the factor matrices may be
chosen to have orthonormal columns, $\mathbf{U}_k^{T}\mathbf{U}_k=\mathbf{I}_{r_k}.$
The mode-$k$ tensor--matrix product is denoted by $\times_k$. Specifically, for
$\ulX\in\mathbb{R}^{p_1\times\cdots\times p_d}$ and
$\mathbf{P}\in\mathbb{R}^{m\times p_k}$, the product
$\ulX\times_k\mathbf{P}$ is a tensor in
$\mathbb{R}^{p_1\times\cdots\times p_{k-1}\times m\times p_{k+1}\times\cdots\times p_d}$,
obtained by multiplying each mode-$k$ fiber of $\ulX$ by $\mathbf{P}$.

The set of tensors with fixed multilinear rank $\mathbf{r}$ is denoted by
\begin{equation}
\mathcal{M}_{\mathbf{r}}
=
\left\{
\ulX\in\mathbb{R}^{p_1\times\cdots\times p_d}
:
\operatorname{rank}(\mathbf{X}_{(k)})=r_k,
\quad k=1,\ldots,d
\right\}.
\end{equation}
For admissible multilinear rank tuple $\textbf{r}$, $\mathcal{M}_{\mathbf{r}}$ forms a smooth
embedded manifold of the ambient tensor space with intrinsic dimension
\citep{KB09}
\begin{equation}\label{eq:dim_man}
\dim(\mathcal{M}_{\mathbf{r}})
=
\prod_{k=1}^{d} r_k
+
\sum_{k=1}^{d}(p_k-r_k)r_k.
\end{equation}
Thus, when $r_k\ll p_k$, the fixed-rank manifold occupies a substantially
lower-dimensional subset of the ambient tensor space.

Given vectors
$\mathbf{b}_k\in\mathbb{R}^{p_k}$,
$k=1,\ldots,d$, their outer product
\[
\mathbf{b}_1\circ\mathbf{b}_2\circ\cdots\circ\mathbf{b}_d
\]
defines a rank-one tensor in
$\mathbb{R}^{p_1\times\cdots\times p_d}$.

\subsection{Low-Rank Tensor Manifold Model for Velocimetry}

The reconstruction begins with the ambient representation of the velocity field in a high-dimensional tensor-product basis,
\begin{equation}\label{eq:amb_model}
v(\xi_1,\ldots,\xi_d;\ultheta )
=
\sum_{i_1=1}^{p_1}
\cdots
\sum_{i_d=1}^{p_d}
\theta_{i_1\cdots i_d}\,
\phi^{(1)}_{i_1}(\xi_1)\cdots\phi^{(d)}_{i_d}(\xi_d),
\end{equation}
where $\ultheta=[\theta_{i_1\cdots i_d}] \in \mathbb{R}^{p_1\times\cdots\times p_d}$ is the coefficient tensor and $\boldsymbol{\xi}=(\xi_1,\ldots,\xi_d)$ denotes a collection of coordinates, each of which may represent either a single physical variable or an arbitrary group of variables. In the higher-order tensor formulation adopted here, each coordinate corresponds to a single physical variable, for example $(\xi_1,\xi_2,\xi_3,\xi_4)=(x_1,x_2,x_3,t)$ for spatiotemporal reconstruction or $(\xi_1,\xi_2,\xi_3)=(x_1,x_2,x_3)$ for single-snapshot reconstruction. More generally, the coordinates may be grouped arbitrarily, allowing the ambient representation to seamlessly interpolate between matrix and higher-order tensor formulations. For example, choosing $\xi_1=(x_1,x_2,x_3)$ and $\xi_2=t$ specializes the representation to a matrix formulation closely related to functional gappy POD, whereas the proposed model retains each physical variable as a separate coordinate to fully exploit multidimensional low-rank structure. The coordinate-wise functions $\{\phi_i^{(j)}\}$ are prescribed basis functions, and a separate ambient representation is constructed for each velocity component. 

In the present work, we use orthonormalized Legendre polynomials as the ambient basis functions. This choice is not intrinsic to the framework, however, and other bases may be employed depending on the problem and coordinate direction, including Fourier bases for periodic coordinates, radial basis functions, and splines.

The corresponding data-misfit functional is
\begin{equation}\label{Eqn:Objective_pol}
\mathcal{L}(\ultheta)
=
\frac{1}{2N}
\sum_{n=1}^{N}
\left(
y^{(n)}
-
v(\boldsymbol{\xi}^{(n)};\ultheta)
\right)^2,
\end{equation}
where $\{\boldsymbol{\xi}^{(n)},y^{(n)}\}_{n=1}^{N}$ are scattered measurements of a single velocity component.

Capturing the broadband multiscale dynamics of turbulent flows typically requires a large number of basis functions, corresponding to an ambient dimension $P=\prod_{k=1}^{d}p_k$. When the number of measurements satisfies $N<P$, the ambient reconstruction problem is underdetermined, admitting infinitely many coefficient tensors that interpolate the observations equally well while producing substantially different flow reconstructions away from the measurement locations. The ambient representation must therefore be restricted to a lower-dimensional model class that regularizes the inverse problem while preserving the essential structure of turbulent flows.

We therefore restrict the coefficient tensor $\ultheta$ to the fixed-rank Tucker manifold $\mathcal{M}_{\mathbf r}$ defined in Section~2(a). Because the intrinsic dimension of $\mathcal{M}_{\mathbf r}$ is typically far smaller than the ambient dimension $P$, the manifold provides a low-dimensional geometric prior that substantially restricts the admissible solution space while retaining an expressive representation of the flow.

The reconstruction is therefore obtained by minimizing the data-misfit functional over the manifold,
\begin{equation}\label{Eqn:Objective_constrained}
\ulhattheta
=
\underset{\ultheta \in \mathcal{M}_{\mathbf r}}{\operatorname{arg\,min}}
\; \mathcal{L}(\ultheta).
\end{equation}

Model complexity is selected adaptively during optimization by increasing the multilinear rank and basis complexity only when the resulting reduction in data misfit exceeds a prescribed threshold. The representation therefore grows only when supported by the measurements, reducing manual model selection. Across all cases considered, this adaptive strategy yielded stable and accurate reconstructions.

\begin{figure}[htpb]
    \centering
    \includegraphics[width=0.8\textwidth]{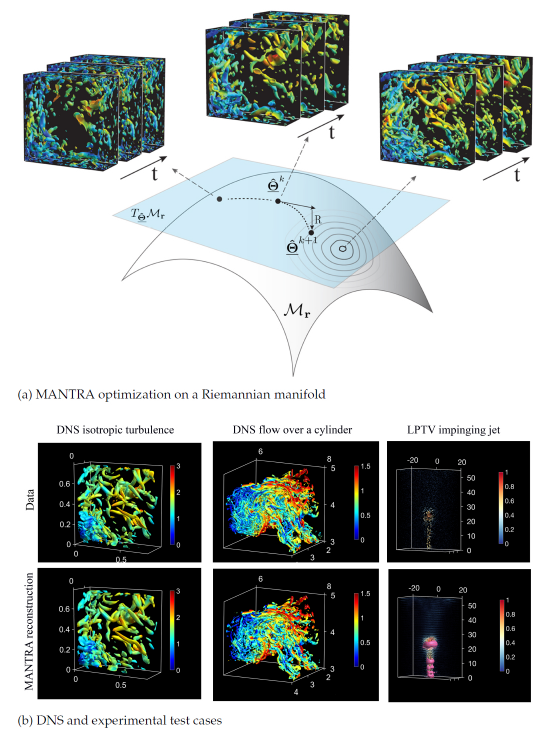}
    \caption{{Manifold-based Tucker low-rank tensor reconstruction of velocity fields via Riemannian optimization.} (a) Schematic of optimization on a low-rank Tucker manifold for a four-dimensional tensor. The iso-surfaces illustrate the refinement of flow structures over iterations. (b) Validation cases include DNS of isotropic turbulence, flow over a cylinder, and an impinging jet experiment. The iso-surfaces represent vorticity magnitude and the color bars indicate velocity magnitude.}
    \label{fig:MANTRA_optimization_and _cases}
\end{figure}

Figure~\ref{fig:MANTRA_optimization_and _cases} summarizes the reconstruction framework and validation cases. Figure ~\ref{fig:MANTRA_optimization_and _cases}a illustrates optimization on the Tucker manifold as the iterative refinement of three-dimensional flow structures over a four-dimensional spatiotemporal window. Performance is assessed using DNS of forced isotropic turbulence and turbulent flow over a cylinder, together with experimental particle-tracking velocimetry data from an impinging jet. Figure~\ref{fig:MANTRA_optimization_and _cases}b compares the corresponding reference measurements or fields (top row) with the MANTRA reconstructions (bottom row).


\subsection{Adaptive Basis and Manifold Geometry}

For the Tucker representation introduced in Section~2(a), the factor matrices have orthonormal columns and therefore satisfy
\begin{equation}
\mathbf{U}_k\in\mathcal{S}(p_k,r_k)
=
\left\{
\mathbf{U}\in\mathbb{R}^{p_k\times r_k}
:
\mathbf{U}^{\mathsf T}\mathbf{U}
=
\mathbf{I}_{r_k}
\right\},
\end{equation}
where $\mathcal{S}(p_k,r_k)$ denotes the Stiefel manifold.

Substitution of the Tucker representation into the ambient representation
induces the coordinate-wise functions
\begin{equation}\label{eq:adaptive_coordinate_basis}
\psi_{\alpha}^{(k)}(\xi_k)
=
\sum_{i=1}^{p_k}
(\mathbf{U}_k)_{i\alpha}
\phi_i^{(k)}(\xi_k),
\qquad
\alpha=1,\ldots,r_k,
\end{equation}
so that the velocity field can be written as
\begin{equation}\label{eq:functional_tucker}
v(\xi_1,\ldots,\xi_d;\ultheta)
=
\sum_{\alpha_1=1}^{r_1}
\cdots
\sum_{\alpha_d=1}^{r_d}
g_{\alpha_1\cdots\alpha_d}
\prod_{k=1}^{d}
\psi_{\alpha_k}^{(k)}(\xi_k).
\end{equation}
Although the ambient functions
$\{\phi_i^{(k)}\}_{i=1}^{p_k}$ are prescribed, the functions
$\{\psi_{\alpha}^{(k)}\}_{\alpha=1}^{r_k}$ are inferred from the
measurements through the factor matrices. Moreover,  $\mathbf{\underline{G}}=[g_{\alpha_1\cdots \alpha_d}] \in \mathbb{R}^{r_1\times\cdots\times r_d}$ is the low-dimensional tensor core, representing the coefficient of the data adaptive bases $\{\psi_{\alpha}^{(k)}\}_{\alpha=1}^{r_k}$. 

Beyond dimensionality reduction, the manifold constraint therefore admits a direct statistical interpretation. As shown in Appendix~\ref{App: Coordinate-wise correlation}, the low-rank structure induces adaptive subspaces along each coordinate direction. For the reconstructed field, the Tucker factor space is the nonzero eigenspace/range of the corresponding coordinate-wise correlation operator formed by averaging over the remaining coordinates. The formulation therefore extends the correlation-operator viewpoint underlying POD from a single global modal subspace to a collection of coordinate-wise subspaces.

Each stage of the
rank-adaptive procedure performs optimization on a manifold with fixed
$\mathbf r$; rank updates transfer the iterate between manifolds of
different multilinear ranks.

An important aspect of the framework is its geometric structure. The fixed-rank Tucker manifold is equipped with well-defined tangent spaces, differential structure, and an associated metric. This geometric viewpoint enables principled linearization, projection, and optimization directly on the manifold. Interestingly, the framework is closely related to dynamical low-rank approximation \citep{KL10,uschmajew2020geometric,GB24}, where evolution equations are solved directly on low-rank manifolds. In the present work, however, the manifold is not used for time integration of PDEs, but rather as a physically structured constraint for inverse reconstruction.


\subsection{Connection to Turbulence Two-Point Correlations Operators}\label{sec:Tucker_flow_structure}

The Tucker representation is closely connected to coordinate-wise
two-point correlation operators. Let
$\mu=\mu_1\otimes\cdots\otimes\mu_d$ be a prescribed product measure, and
assume that the ambient basis functions are orthonormal with respect to
the corresponding coordinate measures. For
$k\in\{1,\ldots,d\}$, define the marginalized correlation kernel
\begin{equation}\label{eq:coordinate_correlation}
C_k(\xi_k,\xi_k')
=
\int_{\Omega_{-k}}
v(\xi_k,\boldsymbol{\xi}_{-k})\,
v(\xi_k',\boldsymbol{\xi}_{-k})\,
\mathrm{d}\mu_{-k}(\boldsymbol{\xi}_{-k}),
\end{equation}
where
$\boldsymbol{\xi}_{-k}$ denotes all coordinates except $\xi_k$ and
$\mu_{-k}=\bigotimes_{j\neq k}\mu_j$.
For square-integrable $v$, this kernel defines a positive-semidefinite,
self-adjoint compact operator on $L^2(\Omega_k,\mu_k)$ and therefore
admits an orthonormal eigendecomposition.

For a field whose coefficient tensor has the Tucker representation in
Eq.~\eqref{eq:Tucker}, let
\begin{equation}
\bm{\Phi}^{(k)}(\xi_k)
=
\begin{bmatrix}
\phi_1^{(k)}(\xi_k) & \cdots & \phi_{p_k}^{(k)}(\xi_k)
\end{bmatrix},
\qquad
\boldsymbol{\Psi}^{(k)}(\xi_k)
=
\boldsymbol{\Phi}^{(k)}(\xi_k)\mathbf U_k .
\end{equation}
Using the orthonormality of the ambient bases and factor matrices,
Eq.~\eqref{eq:coordinate_correlation} becomes
\begin{equation}\label{eq:correlation_tucker}
C_k(\xi_k,\xi_k')
=
\boldsymbol{\Psi}^{(k)}(\xi_k)\,
\mathbf B_k\,
\boldsymbol{\Psi}^{(k)}(\xi_k')^{\mathsf T},
\qquad
\mathbf B_k
=
\mathbf{\underline G}_{(k)}
\mathbf{\underline G}_{(k)}^{\mathsf T},
\end{equation}
where $\mathbf{\underline G}_{(k)}$ denotes the mode-$k$ unfolding of
the core tensor. Let
$\mathbf B_k=\mathbf Q_k\boldsymbol{\Lambda}_k\mathbf Q_k^{\mathsf T}$ and
define
$\widetilde{\boldsymbol{\Psi}}^{(k)}
=\boldsymbol{\Psi}^{(k)}\mathbf Q_k$.
Then
\begin{equation}\label{eq:correlation_eigendecomposition}
C_k(\xi_k,\xi_k')
=
\widetilde{\boldsymbol{\Psi}}^{(k)}(\xi_k)\,
\boldsymbol{\Lambda}_k\,
\widetilde{\boldsymbol{\Psi}}^{(k)}(\xi_k')^{\mathsf T}.
\end{equation}
Thus, the factor space
$\operatorname{span}(\mathbf\Psi^{(k)})$ is the eigenspace of the
coordinate-wise correlation operator associated with the reconstructed
field. The diagonal entries of $\mathbf\Lambda_k$ are the squared
mode-$k$ singular values ($\mathbf{\Lambda_k=}\mathbf{\Sigma}_k^2$). When the Tucker representation is expressed
in HOSVD form,
$\mathbf{\underline G}_{(k)}
\mathbf{\underline G}_{(k)}^{\mathsf T}$
is diagonal, and the factor-induced functions
$\boldsymbol{\Psi}^{(k)}$ are themselves the correlation eigenfunctions.
Further details are provided in Appendix~\ref{App: Coordinate-wise correlation}.

This relationship holds exactly for the reconstructed field. Tucker
structure alone does not guarantee that these functions coincide with
the leading eigenfunctions of the unknown full field. The agreement
with the full-data DNS HOSVD observed in
figure~\ref{fig:Adaptive_bases_and_correlation_p010} therefore demonstrates an
additional and nontrivial property of the sparse-data reconstruction.
\subsection{Riemannian Optimization}

Although the data-misfit functional is quadratic in the ambient
coefficient tensor, its restriction to the fixed-rank Tucker manifold
is nonconvex. We solve the constrained optimization problem, given by Eq.~\eqref{Eqn:Objective_constrained},  using a
nonlinear Riemannian conjugate-gradient (NLRCG) algorithm adapted from
the preconditioned Riemannian framework presented in
\cite{kasai2016low}. The optimization variables are the Tucker
parameters
\[
x=(\mathbf U_1,\ldots,\mathbf U_d,\mathbf{\underline G}),
\]
with each factor matrix constrained to its corresponding Stiefel
manifold.

To account for the different scalings of the core and factor variables,
the product space is endowed with the metric
\begin{equation}\label{eq:riemannian_metric}
g_x(\xi_x,\eta_x)
=
\sum_{i=1}^{d}
\operatorname{tr}
\left(
\xi_{\mathbf U_i}^{\mathsf T}
\eta_{\mathbf U_i}
\mathbf{\underline G}_{(i)}
\mathbf{\underline G}_{(i)}^{\mathsf T}
\right)
+
\left\langle
\xi_{\mathbf{\underline G}},
\eta_{\mathbf{\underline G}}
\right\rangle_F ,
\end{equation}
where $\mathbf{\underline G}_{(i)}$ denotes the mode-$i$ unfolding of
the core. This metric incorporates a block-diagonal approximation of
the Hessian and acts as a geometric preconditioner.

At iteration $k$, the Euclidean derivatives with respect to the core
and factor matrices are converted into the Riemannian gradient and
projected onto the horizontal tangent space. The conjugate search
direction is then formed as
\begin{equation}\label{eq:nlrcg_direction}
\eta_k
=
-\operatorname{grad}\mathcal L(x_k)
+
\beta_k
\mathcal T_{x_{k-1}\rightarrow x_k}(\eta_{k-1}),
\end{equation}
where $\mathcal T$ transports the previous search direction to the
current tangent space and $\beta_k$ is the conjugacy parameter. The
iterate is updated according to
\begin{equation}\label{eq:nlrcg_update}
x_{k+1}
=
\mathcal R_{x_k}(\alpha_k\eta_k),
\end{equation}
where the step length $\alpha_k$ is selected by an Armijo backtracking
line search. The retraction updates the core in its Euclidean space and
restores orthonormality of the factor matrices using an SVD-based
retraction. Model evaluation, gradient expressions, tangent-space
projection, vector transport, retraction, and line search are detailed
in Appendix~\ref{App:Model and Riemannian Optimization}.

The manifold geometry is related to that used in low-rank tensor
completion~\cite{kressner2014low,kasai2016low}, but the measurement
operator is different. Tensor completion observes selected entries of
a discrete tensor, whereas each measurement here is a linear
functional of the coefficient tensor obtained by evaluating the
prescribed basis functions at an arbitrary spatiotemporal coordinate.
The resulting optimization therefore operates directly on scattered
measurements and produces a continuous field without constructing a
full gridded data tensor.


\subsection{Model Adaptivity}\label{Sect.Adaptivity}

Model complexity is controlled independently through the number of
ambient basis functions,
$\mathbf p=(p_1,\ldots,p_d)$, and the Tucker multilinear rank,
$\mathbf r=(r_1,\ldots,r_d)$, with $r_k\leq p_k$ for each mode. The basis
sizes determine the resolution of the ambient function space, whereas
the multilinear ranks determine the dimension of the coordinate-wise
subspaces retained by the low-rank representation. Allowing both
quantities to vary by mode enables the model to account for anisotropic
complexity across spatial and temporal coordinates.

The basis sizes and multilinear ranks are selected using a greedy
mode-wise procedure. The algorithm is initialized with small values of
$\mathbf p$ and $\mathbf r$, with $p_k>r_k$ in each mode. After
optimizing the model at the current complexity, candidate models are
formed by increasing either $p_k$ or $r_k$ in one coordinate direction
at a time. For a candidate update from model $i$ to model $i+1$, the
relative reduction in the optimized loss is defined as
\begin{equation}\label{eq:adaptive_loss_reduction}
\rho
=
\frac{
\mathcal L(\ulhattheta^{\,i})
-
\mathcal L(\ulhattheta^{\,i+1})
}{
\mathcal L(\ulhattheta^{\,i})
}.
\end{equation}
The proposed increase is accepted when
$\rho>\tau_{\mathrm{adapt}}$, where
$\tau_{\mathrm{adapt}}$ is a prescribed tolerance; otherwise, the
corresponding parameter is left unchanged. The accepted model is then
used to initialize the next optimization. This process is repeated
until no increase in any basis size or multilinear rank produces a
relative loss reduction exceeding the tolerance.

The procedure extends the rank-adaptive principle used in low-rank
tensor completion~\cite{kressner2014low} by adapting both the
multilinear rank and the ambient basis resolution independently in each
mode. Because the reconstruction problems considered here are strongly
data limited, the acceptance criterion is evaluated using all available
measurements rather than withholding a separate validation subset.
Although cross-validation or noise-aware information criteria could
also be used, reserving measurements for validation can substantially
reduce the information available for reconstruction in the sparsest
regimes.
\vspace{4cm}
\section{Results}\label{Sect.Results}

\subsection{Benchmarking with Direct Numerical Simulation Data}

Benchmarking is performed using synthetic particle tracks generated from direct numerical simulation (DNS) of two flow configurations: forced isotropic turbulence and turbulent flow over a cylinder. The tracks emulate LPTV measurements by sampling velocity at sparse, irregularly distributed particle locations that evolve in time. Details of the DNS datasets and synthetic particle-track generation are provided in Appendix~\ref{App:DNS Validations}.

For forced isotropic turbulence, we construct two test datasets with particle densities of $0.01$ and $0.04$ particles per voxel, corresponding to approximately $8{,}000$ and $32{,}000$ particles per frame, respectively. These densities are representative of the sparse seeding commonly encountered in volumetric LPTV experiments and correspond to spatial sampling fractions of $1\%$ and $4\%$ relative to the number of DNS grid points per frame. Reconstruction is performed over a $40$-frame observation window. For quantitative evaluation and visualization only, the reconstructed field is subsequently sampled on the DNS Cartesian grid of size $128 \times 128 \times 50 \times 40$.

The second benchmark complements homogeneous isotropic turbulence with an inhomogeneous, wall-bounded flow over a cylinder. In addition to boundary-layer and wake dynamics, the embedded solid boundary introduces a geometry that is not naturally aligned with the tensor-product representation. The synthetic dataset contains approximately $200{,}000$ particles per frame, corresponding to a spatial sampling fraction of $2\%$. No measurements are available inside the cylinder, and its boundary is not explicitly encoded in the reconstruction; nevertheless, the field is reconstructed throughout the surrounding computational domain. The dataset comprises $21$ frames separated by $\Delta t= 0.00375D/U_\infty$. For quantitative evaluation and visualization only, the reconstructed field is sampled a posteriori on the DNS Cartesian grid of size $309 \times 328 \times 87 \times 21$.

Reconstruction accuracy for both DNS benchmarks is quantified using the normalized mean absolute error,
\begin{equation}\label{eq:nmae}
\varepsilon
=
\frac{1}{N_{\mathrm{eval}}\|v\|_{\infty}}
\sum_{i=1}^{N_{\mathrm{eval}}}
\left|v_i-\hat{v}_i\right|,
\end{equation}
where $v_i$ and $\hat{v}_i$ denote the DNS reference and reconstructed values, respectively, at the $N_{\mathrm{eval}}$ points of the Cartesian evaluation grid. The error is computed separately for each velocity component.
\begin{table*}[htb]
\centering
\caption{Component-wise error averaged over time for isotropic turbulence and flow over a cylinder under no noise, 10\% noise and 20\% noise (relative to $|\mathbf{u}|_{max}$).
}
\label{tab:avg_error_noise}
\renewcommand{\arraystretch}{1.15}
\setlength{\tabcolsep}{7pt}

\begin{tabular}{lcc ccc}
\toprule
\textbf{Case} & \textbf{Spatial data density} & \textbf{Noise level}
& ${{\varepsilon}_u}$ & ${{\varepsilon}_v}$ & ${{\varepsilon}_w}$ \\
\midrule

\multirow{6}{*}{Isotropic turbulence}
& \multirow{3}{*}{1\%}
  & No noise               & 0.032  & 0.025 & 0.026 \\
& & $\sigma=0.1|\mathbf{u}|_{max}$  & 0.038  & 0.031 & 0.035 \\
& & $\sigma=0.2|\mathbf{u}|_{max}$  & 0.057  & 0.047 & 0.054 \\
\cmidrule(lr){2-6}
& \multirow{3}{*}{4\%}
  & No noise               & 0.022 & 0.016  & 0.015 \\
& & $\sigma=0.1|\mathbf{u}|_{max}$  & 0.030 & 0.018  & 0.027 \\
& & $\sigma=0.2|\mathbf{u}|_{max}$  & 0.037 & 0.023  & 0.035 \\
\midrule

\multirow{3}{*}{Flow over cylinder}
& \multirow{3}{*}{2\%}
  & No noise               & 0.027 & 0.021 & 0.021  \\
& & $\sigma=0.1|\mathbf{u}|_{max}$  & 0.042 & 0.031 & 0.028 \\
& & $\sigma=0.2|\mathbf{u}|_{max}$  & 0.043 & 0.035 & 0.031 \\

\bottomrule
\end{tabular}
\end{table*}
To assess robustness to measurement noise, we add independent zero-mean Gaussian noise with standard deviations of $10\%$ and $20\%$ of the peak velocity magnitude, producing substantial relative contamination in low-velocity regions. Reconstructions are performed for both DNS benchmarks using noise-free and noise-contaminated measurements. Table~\ref{tab:avg_error_noise} reports the component-wise errors, averaged over time and normalized by the peak velocity magnitude. Without added noise, the errors range from $1.5\%$ to $3.2\%$. Even when the noise standard deviation reaches $20\%$ of the peak velocity, all reconstruction errors remain below $6\%$, indicating that the manifold constraint has an implicit denoising effect in addition to regularizing the undersampled inverse problem. For isotropic turbulence, increasing the spatial sampling fraction from $1\%$ to $4\%$ consistently reduces the error at every noise level. The comparable accuracy obtained for flow over a cylinder further demonstrates robustness across distinct turbulent regimes, including an embedded geometry that is not explicitly encoded in the reconstruction.

\begin{table*}[htb]
\centering
\caption{Component-wise error averaged over time for isotropic turbulence for different reconstruction techniques.
}
\label{tab:avg_error_method}
\renewcommand{\arraystretch}{1.15}
\setlength{\tabcolsep}{7pt}

\begin{tabular}{lcc ccc}
\toprule
\textbf{} & \textbf{4D regression} & \textbf{4D regression }
& \textbf{4D regression} & \textbf{VIC\#} & \textbf{MANTRA} \\
& \textbf{Minimum norm}& \textbf{Ridge reg.}& \textbf{Lasso reg.}& \textbf{Physics reg.} &\\
\midrule

$\varepsilon_u$ &  0.208  & 0.134  & 0.103  & 0.056 & 0.032 \\
$\varepsilon_v$ &  0.207  & 0.121  & 0.138  & 0.047 & 0.025 \\
$\varepsilon_w$ &  0.163  & 0.141  & 0.132  & 0.054 & 0.026 \\

\bottomrule
\end{tabular}
\end{table*}
\begin{figure}[ht]
    \centering
    \includegraphics[width=1\textwidth]{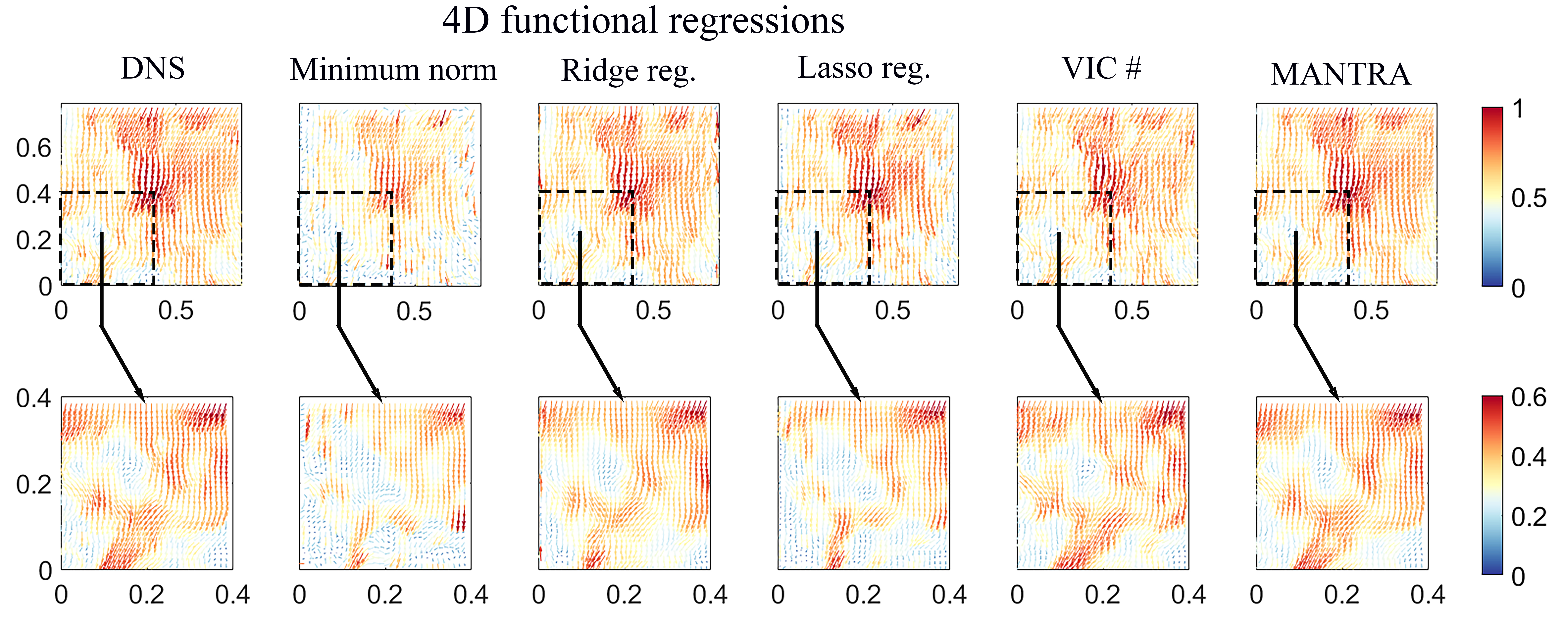}
    \caption{{Instantaneous snapshot of reconstructions performed by different techniques.}
     Vector fields on a $x_1-x_2$ plane (top) and a zoomed region (bottom) comparing reconstruction by 4D functional regression with different regularization, VIC\#, and MANTRA against the DNS. At 1\% spatial data density, adding regularization to functional regression enhances reconstruction quality, MANTRA reconstruction captures more details compared to VIC\#. 
}
    \label{fig:Method Compare}
\end{figure}

The performance of MANTRA is further assessed against baseline reconstruction methods, including the unconstrained 4D functional representation fitted using minimum-norm, Ridge, and Lasso regularization, together with VIC\#, a physics-based data-assimilation method implemented in DaVis 11.0 by LaVision~\cite{jeon2022fine}.  In the Lasso case, the coefficient vector is obtained by solving a $l_1$-regularized least-squares problem using the fast iterative shrinkage-thresholding algorithm~\cite{beck2009fast}. In this algorithm, at each iteration, a gradient step on the data-misfit is followed by a soft-thresholding, promoting sparsity of the coefficients. All methods are evaluated using synthetic particle tracks from isotropic turbulence at a density of $0.01$ particles per voxel. The comparison is particularly stringent because VIC\# assimilates both velocity and acceleration information and enforces physical constraints, whereas MANTRA uses velocity observations alone and requires no governing equations. 
Table~\ref{tab:avg_error_method} reports the component-wise reconstruction errors averaged over time. 
\begin{figure}[ht]
\centering
\includegraphics[width=1\textwidth]{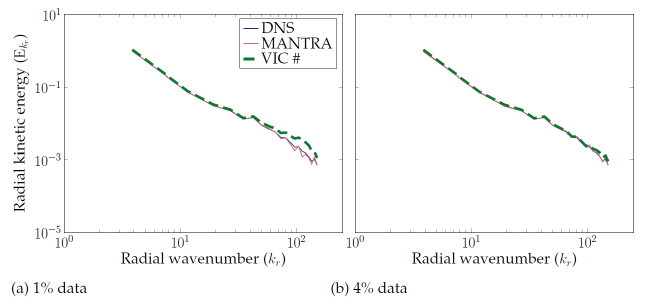}
    \caption{{Radial kinetic energy spectra of the isotropic turbulence computed from the DNS data, VIC\#, and MANTRA.}
    (a) Reconstruction using particle tracks with 1\% spatial data density.
    (b) Reconstruction using particle tracks with 4\% spatial data density.}
    \label{fig:Spectra compare JHU}
\end{figure}
\begin{figure}[H]
\centering
\includegraphics[width=1\textwidth]{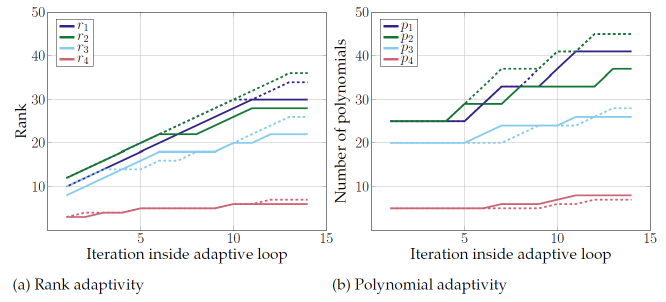}
    \caption{{Change of parameters inside the adaptive loop for the reconstruction of isotropic turbulence using synthetic particle tracks with 1\% (solid lines) and 4\% (dashed lines) spatial data density.} The left and right panels show the change of rank and the number of polynomials, respectively. 
    }
    \label{fig:Polynomial and rank history}
\end{figure}
\begin{table}[htb]
\centering
\caption{Ambient dimension, dimension of the manifold, and training data for MANTRA reconstruction of isotropic turbulence.
}
\label{tab:dimensions_isotropic}
\renewcommand{\arraystretch}{1.15}
\setlength{\tabcolsep}{7pt}

\begin{tabular}{c ccc}
\toprule
\textbf{Spatial data density} & \textbf{Training data} & \textbf{Ambient}
& \textbf{Manifold}\\

\midrule

1\% &  305594  & 315536  & 111562  \\
4\% &  1280365 & 396900  & 235518  \\

\bottomrule
\end{tabular}
\end{table}
The minimum-norm solution yields errors of $16.3\%$--$20.8\%$, demonstrating the instability of the unconstrained ambient representation under sparse sampling. Ridge and Lasso regularization reduce the component-averaged error to approximately $13.2\%$ and $12.4\%$, respectively, while VIC\# lowers it to $5.2\%$. Despite using less input information and no explicit physics, MANTRA achieves the lowest error for every velocity component, with a component-averaged error of $2.8\%$, representing an approximately $47\%$ reduction relative to VIC\#. 
Figure~\ref{fig:Method Compare} shows that this improvement is accompanied by more faithful recovery of local and fine-scale flow structures. The radial kinetic-energy spectra in figure~\ref{fig:Spectra compare JHU} further show that MANTRA more closely reproduces the multiscale energy distribution, particularly at high wavenumbers under the $1\%$ sampling condition. Figure~\ref{fig:Polynomial and rank history} shows the adaptive change of rank and the number of polynomial for these cases. The ambient dimension, dimension of the manifold and the available training data corresponding to these cases are listed in Table~\ref{tab:dimensions_isotropic}

\subsection{Experimental Test Case}

The framework is next applied to experimental LPTV measurements of an oscillating impinging jet, providing a distinct test from the synthetic DNS benchmarks. Unlike homogeneous isotropic turbulence and the cylinder wake, this flow is strongly localized, inhomogeneous, and phase dependent, with dynamics characterized by an oscillating jet core and propagating vortex rings. The measurements were acquired using a four-camera system and processed using the \emph{Shake-the-Box} technique~\cite{schanz2016shake}, implemented in DaVis 11.0 by LaVision~\cite{lavisionShaketheBox}. Details of the experimental configuration, data acquisition, and processing are provided in~\cite{sakib2022pressure}.

\begin{figure}[htb]
    \centering
    \includegraphics[width=0.9\textwidth]{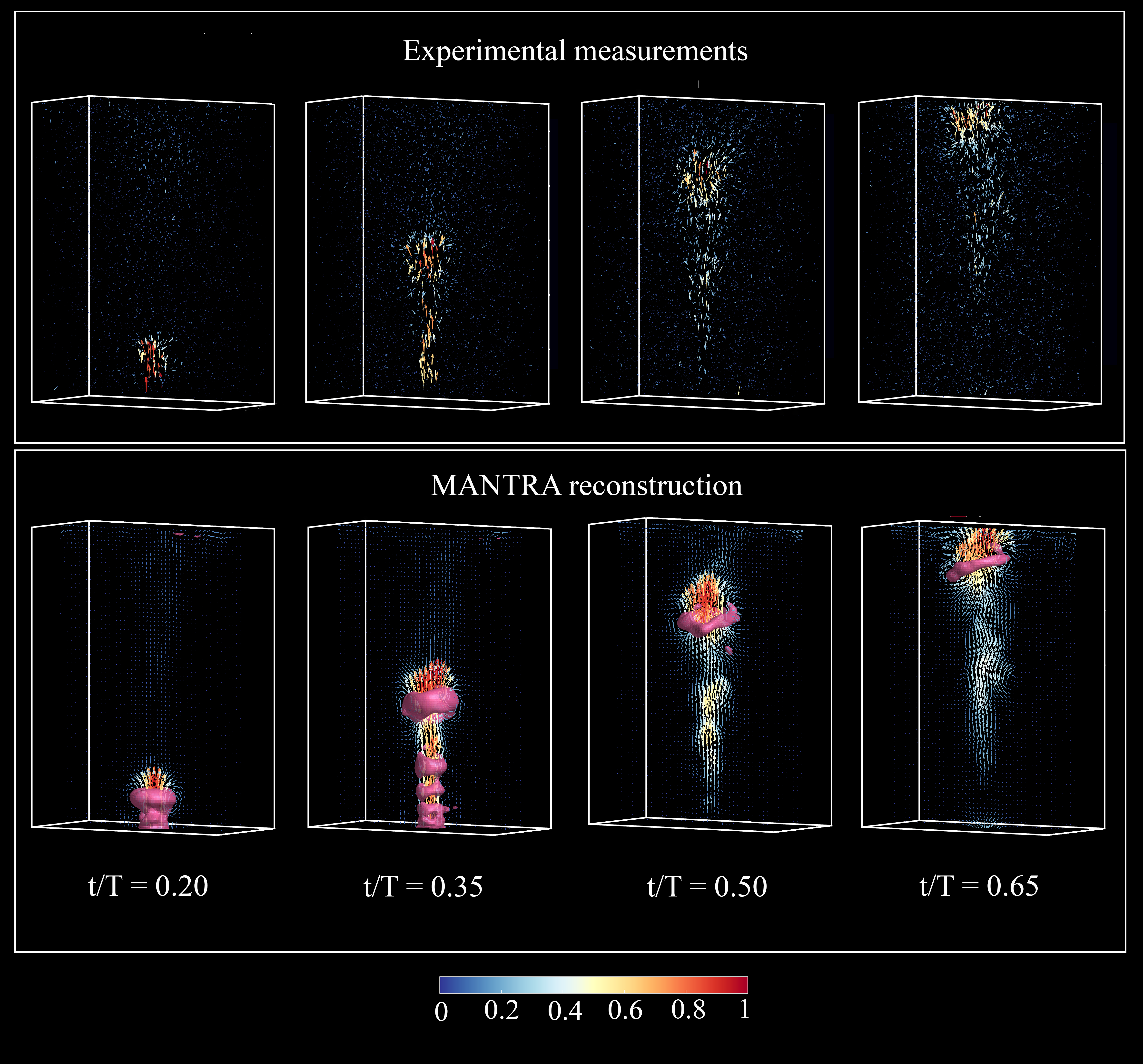}
    \caption{{Reconstruction of an oscillating impinging jet from experimental data}. The top panels show instantaneous snapshots of the scattered data at different phases of the jet. The bottom panels show the reconstructed vector field in the domain along with the iso-surface of vorticity magnitude (pink). The vector colors indicate normalized velocity magnitude.}
    \label{fig:jet_phases}
\end{figure}

The dataset contains $400$ frames spanning one complete oscillation cycle, while each reconstruction uses only a $20$-frame spatiotemporal window. The resulting continuous velocity field is evaluated on a regular grid for visualization and the computation of derived quantities. Figure~\ref{fig:jet_phases} shows the evolution of the jet at several phases of the cycle. The top row contains the instantaneous scattered velocity measurements, and the bottom row shows the corresponding MANTRA reconstruction together with iso-surfaces of vorticity magnitude. The reconstructed fields recover the evolving jet core and the upward propagation of coherent vortex rings, revealing time-dependent structures that are not readily apparent from the scattered measurements alone. This experimental demonstration shows that the same manifold formulation applies directly to real particle-tracking data and to a flow regime qualitatively different from the DNS benchmarks.

\subsection{Model Characteristics}

The flexibility of the ambient representation permits a direct assessment of how coordinate grouping affects reconstruction quality. Three specializations are compared: (i) a matrix formulation, referred to as functional gappy POD, with $d=2$, $\xi_1=(x_1,x_2,x_3)$, and $\xi_2=t$; (ii) a spatial tensor formulation with $d=3$ and $(\xi_1,\xi_2,\xi_3)=(x_1,x_2,x_3)$; and (iii) the proposed spatiotemporal tensor formulation with $d=4$ and $(\xi_1,\xi_2,\xi_3,\xi_4)=(x_1,x_2,x_3,t)$. Two sparse datasets are constructed by retaining every tenth DNS snapshot and either $1\%$ or $0.5\%$ of the spatial samples within each retained snapshot, corresponding to overall sampling fractions of approximately $10^{-3}$ and $5\times10^{-4}$ of the full spatiotemporal DNS data, respectively. The functional gappy POD and spatiotemporal formulations use identical measurements, thereby isolating the effect of coordinate grouping, whereas the spatial formulation uses measurements from a single snapshot and serves as the purely spatial counterpart. Model complexity is selected adaptively for the spatial and spatiotemporal formulations. For a controlled comparison, functional gappy POD uses the parameter values selected by the corresponding spatiotemporal reconstruction. Holding these values fixed makes differences between the two formulations more directly attributable to coordinate grouping.

\begin{figure}[p]
    \centering
    \includegraphics[width=0.9\textwidth]{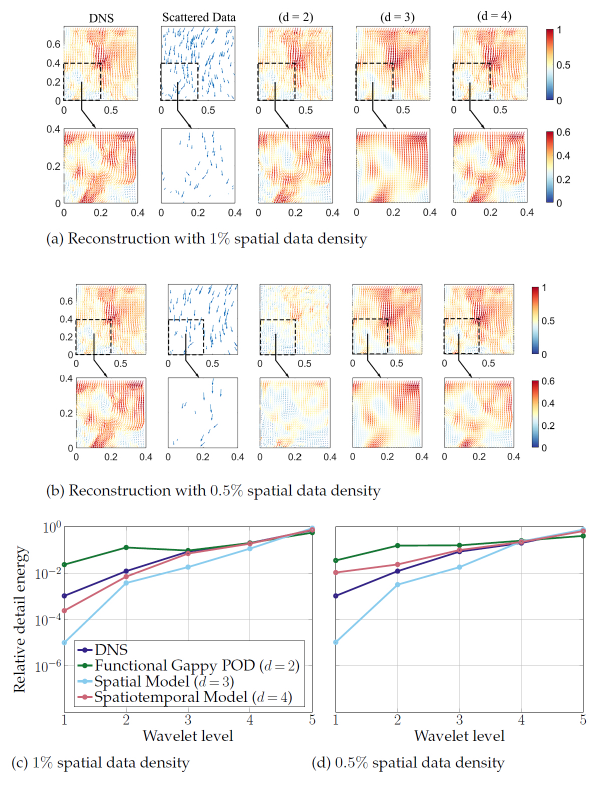}

    \caption{{Effect of coordinate grouping on reconstruction quality.}
Velocity fields on an $x_1$--$x_2$ plane for
(a) $1\%$ and (b) $0.5\%$ spatial sampling. Full fields
are shown above and enlarged regions below, comparing DNS, scattered
observations, and the 2D, 3D, and 4D reconstructions.
(c, d) Corresponding wavelet spectra.
}
    \label{fig:Correlation_effect}
\end{figure}

Figure~\ref{fig:Correlation_effect}a shows that, at a spatial sampling fraction of $1\%$, all three formulations reproduce the large-scale flow organization at the full-field level. Differences become apparent in the enlarged regions, where the spatiotemporal tensor formulation agrees most closely with the DNS and more accurately resolves fine-scale structures. These differences become more pronounced when the spatial sampling fraction is reduced to $0.5\%$ (figure~\ref{fig:Correlation_effect}b). Under this more severe undersampling, functional gappy POD fails to recover some large-scale features, while the spatial tensor formulation produces a smoother field with diminished small-scale content. The spatiotemporal formulation remains closest to the DNS across both scales. The wavelet spectra confirm these observations: the spatiotemporal formulation most closely reproduces the DNS energy distribution at both sampling fractions (figures ~\ref{fig:Correlation_effect}c and ~\ref{fig:Correlation_effect}d), whereas functional gappy POD is inaccurate even for part of the large-scale energy at $0.5\%$ sampling. The size of the training data, ambient dimension, and the dimension of the manifold for the 2D, 3D, and 4D models are listed in table~\ref{tab:MANTRA_model_dimensions_isotropic}

The distinction arises from how the coordinates are organized. Functional gappy POD groups all spatial variables into a single coordinate and can therefore represent global space--time correlations, but it cannot separately exploit low-rank correlations along the individual spatial directions and time. The spatiotemporal tensor formulation preserves these coordinate-wise structures and assigns an adaptive subspace to each coordinate. This organization yields a more data-efficient representation: despite having a substantially smaller manifold dimension than functional gappy POD, the spatiotemporal formulation provides consistently more accurate reconstruction under severe undersampling.
\begin{table}[htb!]
\centering
\caption{Ambient dimension, dimension of the manifold, and training data for different MANTRA models. 
}
\label{tab:MANTRA_model_dimensions_isotropic}
\renewcommand{\arraystretch}{1.15}
\setlength{\tabcolsep}{7pt}

\begin{tabular}{c cccc}
\toprule
 \textbf{Spatial data density} & \textbf{Model}
& \textbf{Training data} & \textbf{Ambient} & \textbf{Manifold} \\
\midrule

 \multirow{3}{*}{1\%}
  & $d = 2$ & 163000  & 205800  & 78408 \\
  & $d = 3$ & 8150    & 196000  & 7056  \\
  & $d = 4$ & 163000  & 205800  & 34608 \\
\midrule
 \multirow{3}{*}{0.5\%}
  & $d = 2$ & 81000 & 90000  & 50408 \\
  & $d = 3$ & 4050  & 18000  & 5746  \\
  & $d = 4$ & 81000 & 90000  & 20722 \\
\bottomrule
\end{tabular}
\end{table}

\begin{figure}[ht]
\centering
\includegraphics[width=1\textwidth]{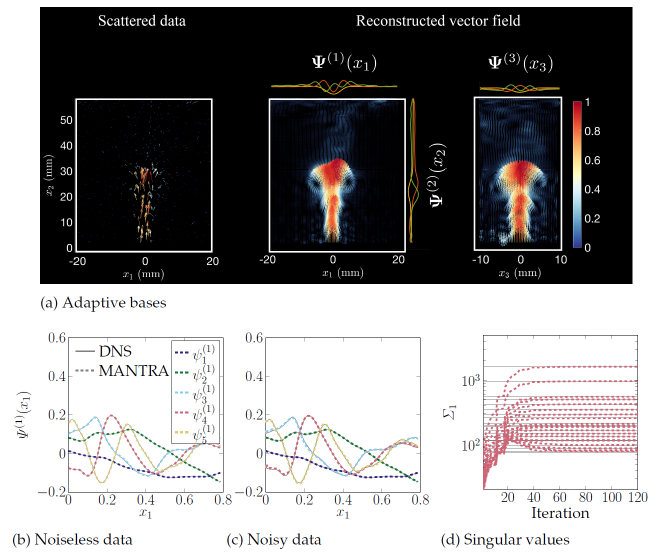}
    \caption{{Data-adaptive bases and recovery of correlation structure from sparse velocimetry data.}
(a) Impinging-jet experiment: scattered velocity vectors in a thin $x_1$--$x_2$ slice
($-0.5~{\rm mm}\leq x_3\leq0.5~{\rm mm}$), reconstructed mid-plane velocity fields, and the first three adaptive basis functions along $x_1$, $x_2$ and $x_3$.
(b--d) Isotropic turbulence: recovery of the $x_1$ correlation structure from sparse measurements.
(b) First five DNS correlation eigenfunctions (solid) and reconstructed eigenfunctions (dashed) without added noise.
(c) Corresponding eigenfunctions with $10\%$ Gaussian noise.
(d) Convergence of the leading $x_1$ singular values toward the DNS values during optimization.}
    \label{fig:Adaptive_bases_and_correlation_p010}
\end{figure}

While we use global 1D bases in each coordinate direction, the low-rank constraint modifies these bases through the factor matrices, enabling the model to adapt to localized flow variability.
These adaptive bases make the mechanism of the model interpretable, i.e., rather than employing a large number of fixed bases to represent the varying scales in the flow, the model concentrates capacity along dominant flow structures. Figure~\ref{fig:Adaptive_bases_and_correlation_p010}a shows the first 3 dominant adaptive bases in each coordinate direction for the impinging jet experiment.
The manifold representation has a direct interpretation in turbulence structure analysis: its coordinate-wise subspaces approximate the dominant eigenspaces of marginalized two-point correlation operators. To assess whether these dominant structures can be inferred from sparse observations, we compute the mode-wise higher-order singular value decomposition (HOSVD)~\cite{KB09} of the MANTRA reconstruction and compare the resulting correlation eigenfunctions with those obtained from the complete DNS field. Figure~\ref{fig:Adaptive_bases_and_correlation_p010}b shows that the first five eigenfunctions along $x_1$ recovered from sparse, noise-free measurements closely match the leading full-data HOSVD eigenfunctions. This agreement persists under $10\%$ measurement noise (figure~\ref{fig:Adaptive_bases_and_correlation_p010}c), and the corresponding leading singular values converge toward the full-data values during optimization (figure~\ref{fig:Adaptive_bases_and_correlation_p010}d). These results demonstrate that MANTRA recovers the leading coordinate-wise singular subspaces of the full flow field from sparse and noisy measurements, rather than an arbitrary basis spanning a low-rank representation. 

The corresponding results for the $4\%$ spatial sampling case are provided in figure~\ref{fig:Adaptive_bases_and_correlation_p040} in Appendix~\ref{App:DNS Validations}.

\section{Conclusion and Perspective}
The broader implication of this work is that low-rank tensor manifolds can serve as effective inductive biases for turbulent-flow reconstruction. Low-rank manifolds have recently been used for on-the-fly reduced-order modeling of turbulent flows by constraining solution evolution and reducing computational complexity \cite{GLD22,JLBC25,GGJP25}. The present results show that the same geometric principle can regularize the inverse problem of recovering turbulent fields from sparse, scattered, and noisy measurements. Restricting the reconstruction to a structured low-dimensional set reduces ambiguity while retaining a continuous representation that adapts to the observations. Tensor manifolds may therefore provide a common geometric framework for reduced-order simulation, data assimilation, and sparse-data reconstruction.

The Tucker manifold was adopted here because it combines an interpretable multilinear structure with a well-defined smooth geometry and mature Riemannian optimization algorithms. Its strong performance across the present test cases shows that Tucker structure is itself an effective inductive bias for turbulent-field reconstruction. At the same time, this does not imply that Tucker is universally optimal for every turbulent flow. More complex tensor-network architectures may represent anisotropy, multiscale coupling, intermittency, coherent structures, and boundary effects more efficiently. Importantly, Tucker need not be viewed as an alternative to such representations; it can also appear as a building block within hierarchical Tucker, multilevel, localized, or hybrid tensor networks. We therefore expect future tensor representations for turbulence to introduce additional connectivity, locality, or multiscale structure tailored to the statistical and dynamical organization of the flow.

The recovery of the leading coordinate-wise HOSVD subspaces from sparse and noisy measurements is particularly significant. It indicates that the manifold constraint can recover dominant statistical structures of the complete flow field from incomplete observations. 

A complete theory connecting this property to the recovery of leading correlation eigenspaces remains open. 

Several extensions follow naturally from this perspective. Physical constraints could be incorporated when reliable governing equations, forcing terms, and boundary data are available, while preserving the ability to operate directly from measurements when they are not. Geometry-aware and localized tensor representations may improve reconstruction near embedded boundaries, sharp gradients, and strongly intermittent regions. More broadly, these results motivate viewing tensor networks as adaptable scientific priors whose structure can be designed jointly with the governing dynamics, the geometry of the flow, and the measurement process.

\section*{Code Availability}
The code for MANTRA implemented in Matlab can be accessed from the following GitHub repository: \href{https://github.com/MsakibPitt/MAntra-Codes}{MANTRA Codes}
\section*{Acknowledgment}{This work was supported by the U.S. Department of Energy (Grant No.  DEAR140449). The authors gratefully acknowledge this support. 
Notice: This report was prepared as an account of work sponsored by an agency of the United States Government. Neither the United States Government nor any agency thereof, nor any of their employees, nor any of their contractors, subcontractors or their employees, makes any warranty, express or implied, or assumes any legal liability or responsibility for the accuracy, completeness, or any third party’s use or the results of such use of any information, apparatus, product, or process disclosed, or represents that its use would not infringe privately owned rights. Reference herein to any specific commercial product, process, or service by trade name, trademark, manufacturer, or otherwise, does not necessarily constitute or imply its endorsement, recommendation, or favoring by the United States Government or any agency thereof or its contractors or subcontractors. The views and opinions of authors expressed herein do not necessarily state or reflect those of the United States Government or any agency thereof.}

\section*{Use of artificial intelligence}

AI tools were used solely for language editing to improve the
readability of the manuscript. All suggested changes were reviewed
and, where appropriate, revised by the authors, who take full
responsibility for the content of the article.

\section*{Appendix}
\appendix
\renewcommand{\theequation}{\thesection\arabic{equation}}
\makeatletter
\@addtoreset{equation}{section}
\makeatother

\section{Coordinate-wise Correlation Operator and Tucker Decomposition}\label{App: Coordinate-wise correlation}

We now show how the Tucker structure is related to coordinate-wise correlation operators. 
We assume that the prescribed basis functions are orthonormal in each coordinate direction:
\[
\int_{\Omega_k}
\phi^{(k)}_{i}(\xi_k)
\phi^{(k)}_{j}(\xi_k)
\,d\xi_k
=
\delta_{ij}.
\]
Considering the first coordinate direction. The coordinate-wise correlation operator is defined by averaging over all remaining coordinates:
\[
C_1(\xi_1,\xi_1')
=
\int_{\Omega_2}\cdots\int_{\Omega_d}
v(\xi_1,\xi_2,\ldots,\xi_d)
v(\xi_1',\xi_2,\ldots,\xi_d)
\,d\xi_2\cdots d\xi_d.
\]
Substituting the ambient expansion gives
\[
\begin{aligned}
C_1(\xi_1,\xi_1')
&=
\int_{\Omega_2}\cdots\int_{\Omega_d}
\left[
\sum_{i_1,\ldots,i_d}
\theta_{i_1\cdots i_d}
\phi^{(1)}_{i_1}(\xi_1)
\prod_{k=2}^d \phi^{(k)}_{i_k}(\xi_k)
\right]
\\
&\qquad\qquad\times
\left[
\sum_{j_1,\ldots,j_d}
\theta_{j_1\cdots j_d}
\phi^{(1)}_{j_1}(\xi_1')
\prod_{k=2}^d \phi^{(k)}_{j_k}(\xi_k)
\right]
\,d\xi_2\cdots d\xi_d .
\end{aligned}
\]
Using orthonormality in the averaged coordinates,
all cross terms vanish unless
\[
i_k=j_k,\qquad k=2,\ldots,d.
\]
Therefore,
\[
C_1(\xi_1,\xi_1')
=
\sum_{i_1=1}^{p_1}
\sum_{j_1=1}^{p_1}
\left[
\sum_{i_2,\ldots,i_d}
\theta_{i_1 i_2\cdots i_d}
\theta_{j_1 i_2\cdots i_d}
\right]
\phi^{(1)}_{i_1}(\xi_1)
\phi^{(1)}_{j_1}(\xi_1').
\]

Let \(\ultheta_{(1)}\in \mathbb{R}^{p_1\times (p_2\cdots p_d)}\) denote the mode-1 unfolding of \(\ultheta\). Then the coefficient matrix appearing in the kernel is
\[
\mathbf K_1
=
\ultheta_{(1)}\ultheta_{(1)}^\top
\in \mathbb{R}^{p_1\times p_1}.
\]
Hence,
\[
C_1(\xi_1,\xi_1')
=
\mathbf{\Phi}^{(1)}(\xi_1)
\mathbf K_1
\mathbf{\Phi}^{(1)}(\xi_1')^\top,
\]
where
\[
\boldsymbol{\Phi}^{(1)}(\xi_1)
=
\begin{bmatrix}
\phi^{(1)}_1(\xi_1)&
\cdots&
\phi^{(1)}_{p_1}(\xi_1)
\end{bmatrix}.
\]
The same argument applies to every coordinate direction \(k\). In general,
\[
C_k(\xi_k,\xi_k')
=
\boldsymbol{\Phi}^{(k)}(\xi_k)
\mathbf K_k
\boldsymbol{\Phi}^{(k)}(\xi_k')^\top,
\qquad
\mathbf K_k
=
\ultheta_{(k)}\ultheta_{(k)}^\top.
\]

We next show how the Tucker representation of the coefficient tensor induces low-dimensional coordinate-wise correlation subspaces. Representing the coefficient tensor $\ultheta$ with a low-rank Tucker factorization of the form

\[
\ultheta
\approx
\ulhattheta
=
\mathbf{\underline{G}}\times_1 \mathbf{U}_1\times_2 \mathbf{U}_2\cdots \times_d \mathbf{U}_d,
\]
with
\[
\mathbf{U}_k^\top \mathbf{U}_k=\mathbf{I}_{r_k}.
\]
The mode-\(k\) unfolding of \(\ulhattheta\) is
\[
\ulhattheta_{(k)}
=
\mathbf{U}_k
\mathbf{\underline{G}}_{(k)}
\left(
\mathbf{U}_d\otimes \cdots \otimes \mathbf{U}_{k+1}
\otimes \mathbf{U}_{k-1}\otimes \cdots \otimes \mathbf{U}_1
\right)^\top,
\]
where \(\mathbf{\underline{G}}_{(k)}\) is the mode-\(k\) unfolding of the core tensor. Therefore,
\[
\begin{aligned}
\mathbf K_k
&\approx
\ulhattheta_{(k)}\ulhattheta_{(k)}^\top
\\
&=
\mathbf{U}_k
\mathbf{\underline{G}}_{(k)}
\left(
\mathbf{U}_d\otimes \cdots \otimes \mathbf{U}_{k+1}
\otimes \mathbf{U}_{k-1}\otimes \cdots \otimes \mathbf{U}_1
\right)^\top
\\
&\qquad \times
\left(
\mathbf{U}_d\otimes \cdots \otimes \mathbf{U}_{k+1}
\otimes \mathbf{U}_{k-1}\otimes \cdots \otimes \mathbf{U}_1
\right)
\mathbf{\underline{G}}_{(k)}^\top
\mathbf{U}_k^\top .
\end{aligned}
\]
Since all factor matrices have orthonormal columns,
\[
\left(
\mathbf{U}_d\otimes \cdots \otimes \mathbf{U}_{k+1}
\otimes \mathbf{U}_{k-1}\otimes \cdots \otimes \mathbf{U}_1
\right)^\top
\left(
\mathbf{U}_d\otimes \cdots \otimes \mathbf{U}_{k+1}
\otimes \mathbf{U}_{k-1}\otimes \cdots \otimes \mathbf{U}_1
\right)
=
\mathbf{I}.
\]
Hence,
\[
 \mathbf K_k
\approx
\mathbf{U}_k
\mathbf{\underline{G}}_{(k)}
\mathbf{\underline{G}}_{(k)}^\top
\mathbf{U}_k^\top.
\]
For Tucker factorization obtained through HOSVD, $\mathbf{\underline{G}}_{(k)}\mathbf{\underline{G}}_{(k)}^\top$ is diagonal where the entries are the squared singular values of the $k$ mode unfolding of $\mathbf{\underline{G}}$. Therefore,
\[
\mathbf K_k
\approx
\mathbf{U}_k
\boldsymbol{\Lambda_k}
\mathbf{U}_k^\top,
\]
where, $\boldsymbol{\Lambda_k}$ is the diagonal matrix containing the eigenvalues of $\mathbf{\underline{G}}_{(k)}\mathbf{\underline{G}}_{(k)}^\top$.
Substituting $K_k$ into the coordinate-wise correlation operator $C_k(\xi_k,\xi_k')$,
\[
\begin{aligned}
    C_k & \approx\boldsymbol{\Phi}_k\left(\xi_k\right)\mathbf{U}_k\boldsymbol{\Lambda}_k
\mathbf{U}_k^\top\boldsymbol{\Phi}_k\left(\xi_k\right)^\top\\
  & = \boldsymbol{\Psi}_k\left(\xi_k\right)\boldsymbol{\Lambda}_k\boldsymbol{\Psi}_k\left(\xi_k\right)^\top,
\end{aligned}
\]
where, $\boldsymbol{\Psi}_k\left(\xi_k\right)=\boldsymbol{\Phi}_k\left(\xi_k\right)\mathbf{U}_k$ are the data-adaptive basis functions.

\section{Model Evaluation and the Elements of Riemannian Optimization}\label{App:Model and Riemannian Optimization}

\subsection{Model Evaluation}
Let $\boldsymbol \xi^{(n)} = [\xi^{(n)}_1,\xi^{(n)}_2, \dots, \xi^{(n)}_d]$ denote the $n$th labeled input coordinates and moreover let us define $\mathbf \Phi_i \in \mathbb R^{N \times p_i}$ such that $\mathbf \Phi_i(n,j) = \phi_{j_i}(\xi^{(n)})$.
Additionally, let us denote $\mathbf b^\top_{i,n} = \mathbf \Phi_i(n,:)$ as the $n$th row of the $\mathbf \Phi_i$ matrix.  

The model evaluated at the $n$th training point is a scalar given by  $v(\boldsymbol \xi^{(n)};\ultheta) \in \mathbb R$, and can be expressed as:
\begin{equation}
    v(\boldsymbol \xi^{(n)};\ultheta) = \ultheta \times_1 \mathbf b^\top_{1,n} \times_2 \mathbf b^\top_{2,n} \dots \times_d \mathbf b^\top_{d,n}.
\end{equation}

The derivative of this scalar with respect to the coefficient tensor ($\ultheta$) is a tensor of the same size as $\ultheta$ as follows
\begin{equation}
    \frac {\partial v(\boldsymbol \xi^{(n)};\ultheta)}{\partial \ultheta} = \mathbf  b_{1,n} \circ \mathbf b_{2,n} \dots  \circ \mathbf b_{d,n}, 
\end{equation}
where $[\circ]$ denotes tensor outer product.

The gradient of $\mathcal{L}({\ultheta})$ with respect to $\ultheta$ is a tensor of the same size as $\ultheta$ as follows:
\begin{equation}
    \nabla_{\ultheta} \mathcal L = \frac{1}{N}\sum_{n=1}^N\left(v({\boldsymbol \xi^{(n)};\ultheta})-y^{(n)}\right) \frac {\partial v(\boldsymbol \xi^{(n)};\ultheta)}{\partial \ultheta}.
\end{equation}
The term $v({\boldsymbol \xi^{(n)};\ultheta}) \dfrac {\partial v(\boldsymbol \xi^{(n)};\ultheta)}{\partial \ultheta}$ can be worked out as follows: 
\begin{equation}
  v({\boldsymbol \xi^{(n)};\ultheta}) \dfrac {\partial v(\boldsymbol \xi^{(n)};\ultheta)}{\partial \ultheta} = 
    {\ultheta} \times_1 (\mathbf  b_{1,n} \mathbf  b^\top_{1,n}) \times_2 (\mathbf  b_{2,n} \mathbf  b^\top_{2,n}) \dots \times _d (\mathbf  b_{d,n} \mathbf  b^\top_{d,n}).  
\end{equation}
Therefore, the gradient of the loss function can be written as:
\begin{equation}\label{eqn:gradien_tensor}
    \nabla_{\ultheta} \mathcal L =\frac{1}{N}\sum_{n=1}^N \left[\mathbf{\ultheta} \times_1 (\mathbf  b_{1,n} \mathbf  b^\top_{1,n}) \times_2 (\mathbf  b_{2,n} \mathbf  b^\top_{2,n}) \dots \times _d (\mathbf  b_{d,n} \mathbf  b^\top_{d,n})-y^{(n)}\big(\mathbf  b_{1,n} \circ \mathbf b_{2,n}  \dots \circ \mathbf b_{d,n} \big) \right].
\end{equation}

\subsection{Riemannian Optimization}
\label{sec: NLRCG}

We endow the low-rank polynomial coefficient tensor model with a Riemannian structure analogous to~\cite{kasai2016low}. The ambient object is the coefficient tensor 
\[
\ultheta\in\mathbb{R}^{p_1\times p_2 \times \dots \times p_d},
\]
and its Tucker approximation is 
\begin{equation}
  \ulhattheta
  = \underline{\mathbf{G}} \times_1 \mathbf{U}_1 \times_2 \mathbf{U}_2 \cdots \times_d \mathbf{U}_d,
  \qquad
  \mathbf{U}_i \in \mathbb{R}^{p_i \times r_i}, \;
  \mathbf{U}_i^\top \mathbf{U}_i = \mathbf{I}_{r_i}.
  \label{eq:tucker-approx}
\end{equation}
The loss function is

\begin{equation}
    \mathcal{L}(\ulhattheta) = \frac{1}{2N} \sum_{n=1}^N
    \bigl\lvert y^{(n)} - v(\boldsymbol \xi^{(n)}; \ulhattheta) \bigr\rvert^2,
\end{equation}
where
\[
  v(\boldsymbol \xi^{(n)}; \ulhattheta)
  =
  \ulhattheta
  \times_1 \mathbf{b}_{1,n}^\top
  \times_2 \mathbf{b}_{2,n}^\top
  \cdots
  \times_d \mathbf{b}_{d,n}^\top.
\]

We optimize over the Tucker parameters

\[
x=(\mathbf{U}_1, \mathbf{U}_2, \dots,\mathbf{U}_d,\underline{\mathbf{G}},)
\]
where each factor matrix $\mathbf{U}_i \in \mathbb{R}^{p_i \times r_i}$ satisfies
$\mathbf{U}_i^\top \mathbf{U}_i = \mathbf{I}_{r_i}$ for $i = 1,\dots,d$, and
$\underline{\mathbf{G}} \in \mathbb{R}^{r_1 \times \cdots \times r_d}$ is the core tensor.
Let $\mathbf{G}_i$ denote the mode-$i$ unfolding of $\underline{\mathbf{G}}$. 

The elements of NLRCG has the following forms:

\paragraph*{\rm\textbf{Euclidean gradient of the objective function.}}
The Euclidean gradient of the objective function has the following form:
\begin{equation}
    \nabla_{\ultheta} \mathcal L =\frac{1}{N}\sum_{n=1}^N \left[\mathbf{\ultheta} \times_1 (\mathbf  b_{1,n} \mathbf  b^\top_{1,n}) \times_2 (\mathbf  b_{2,n} \mathbf  b^\top_{2,n}) \dots \times _d (\mathbf  b_{d,n} \mathbf  b^\top_{d,n})-y^{(n)}\big(\mathbf  b_{1,n} \circ \mathbf b_{2,n}  \dots \circ \mathbf b_{d,n} \big) \right].
\end{equation}

\paragraph*{\rm\textbf{Riemannian metric}}
The Riemannian metric at $x$ acting on tangent vectors 
\[
\xi_x = (\xi_{\mathbf{U}_1}, \cdots \xi_{\mathbf{U}_d},\underline{\xi}_\mathbf{G} )
\qquad
\eta_x = (\eta_{\mathbf{U}_1}, \cdots \eta_{\mathbf{U}_d},\underline{\eta}_\mathbf{G} )
\]
is
\begin{equation}
    g_x(\xi_x, \eta_x)
  =
  \sum_{i=1}^d
  \operatorname{tr}
  \bigl(
    \xi_{\mathbf{U}_i}^\top \eta_{\mathbf{U}_i} \, \mathbf{G}_i \mathbf{G}_i^\top
  \bigr)
  \;+\;
  \langle \xi_\mathbf{G}, \eta_\mathbf{G} \rangle_F.
\end{equation}

\paragraph*{\rm\textbf{Tangent space projection.}}
Given an ambient vector
\[
y=(y_{\mathbf{U}_1}, \dots, y_{\mathbf{U}_d}, \underline{y}_{\mathbf{G}})
\]
the tangent space projection of $y$ at $x$, is given as
\begin{equation}
   \mathcal{T}_{x}\mathcal{M}(y_{\mathbf{U}_i}) := {y_{\mathbf{U}_i}-\mathbf{U_i}}\mathbf{S_{U_i}}(\mathbf{G_i\mathbf{G^\top_i}})^{-1},
   \qquad
   \mathcal{T}_{x}\mathcal{M}(\underline{y}_{\mathbf{G}}):=\underline{y}_{\mathbf{G}},
\end{equation}
where $\mathbf{S_{\mathbf{U}_i}}$ are the solution of the Lyapunov equations of the following form
\[
\mathbf{S_{\mathbf{U}_i}} \mathbf{G_i\mathbf{G^\top_i}} + \mathbf{G_i\mathbf{G^\top_i}} \mathbf{S_{\mathbf{U}_i}} = \mathbf{G_i\mathbf{G^\top_i}}(y^\top_{\mathbf{U}_i} \mathbf{U}_i + \mathbf{U}^\top_i y_{\mathbf{U}_i}) \mathbf{G_i\mathbf{G^\top_i}}, \quad i=1,2,\dots,d.
\]
Since the objective is invariant under the Tucker quotient
transformations, its Riemannian gradient is orthogonal to all
vertical directions. Hence the Riemannian gradient is already
horizontal, and no additional horizontal-space projection is needed.
\paragraph*{\rm\textbf{Retraction.}}

We use an SVD-based orthogonal factor retraction.
For any matrix $A \in \mathbb{R}^{p_i \times r_i}$, define
\[
  u_f(\mathbf{A}) := \mathbf{L} \mathbf{R}^\top,
  \quad\text{where } \mathbf{A} = \mathbf{L} \mathbf{\Sigma} \mathbf{R}^\top
  \text{ is a (thin) SVD.}
\]
Given a point
\[
  x = (\mathbf{U}_1, \dots, \mathbf{U_d}, \mathbf{\underline{G}})
\]
and a tangent vector
\[
  \xi_x = (\xi_{\mathbf{U}_1}, \dots, \xi_{\mathbf{U}_d}, \xi_{\mathbf{\underline{G}}}),
\]
the retraction $R_x : T_x\mathcal{M} \to \mathcal{M}$ for a step size $\alpha > 0$ is defined as
\begin{align*}
  \mathbf{U}_i^{+} &:= u_f\bigl(\mathbf{U}_i + \alpha \,\xi_{\mathbf{U}_i}\bigr),
  \qquad i = 1,\dots,d, \\
  \mathbf{\underline{G}}^{+} &:= \mathbf{\underline{G}} + \alpha \,\xi_\mathbf{\underline{G}}.
\end{align*}
Thus,
\[
  R_x(\alpha \xi_x) = x^{+}
  = (\mathbf{U}_1^{+}, \dots, \mathbf{U}_d^{+}, \mathbf{\underline{G}}^{+}).
\]

\paragraph*{\rm\textbf{Vector transport}}

Nonlinear conjugate gradient requires a vector transport
\(
  T_{x_k \to x_{k+1}} : T_{x_k}\mathcal{M} \to T_{x_{k+1}}\mathcal{M}.
\)
A standard choice, consistent with the above geometry, is the following
projected transport.

At iteration $k$, suppose
\[
  x_k = (\mathbf{U}_1, \dots, \mathbf{U}_d, \mathbf{\underline{G}}),
  \qquad
  \eta_k = (\eta_{\mathbf{U}_1}, \dots, \eta_{\mathbf{U}_d}, \eta_\mathbf{\underline{G}}) \in T_{x_k}\mathcal{M}
\]
is the current search direction, and we step to
\[
  x_{k+1}
  = R_{x_k}(\alpha_k \eta_k)
  = (\mathbf{U}_1^{+}, \dots, \mathbf{U}_d^{+}, \mathbf{\underline{G}}^{+}).
\]
We first regard $\eta_k$ as an ambient vector at the new point:
\[
  \tilde{\eta}
  = (\eta_{\mathbf{U}_1}, \dots, \eta_{\mathbf{U}_d}, \eta_\mathbf{\underline{G}}),
\]
and then project it back to the tangent space $T_{x_{k+1}}\mathcal{M}$ using the
tangent projector.

Let $\mathbf{G}_i^{+}$ be the mode-$i$ unfolding of $\mathbf{\underline{G}}^{+}$. For each $i = 1,\dots,d$,
solve the Lyapunov equation for the symmetric matrix $\mathbf{S}_{\mathbf{U}_i}^{+}$:
\begin{equation}
  \mathbf{S}_{\mathbf{U}_i}^{+} \mathbf{G}_i^{+} \mathbf{G}^{+\top}
  + \mathbf{G}^{+} \mathbf{G}^{+\top} \mathbf{S}_{\mathbf{U}_i}^{+}
  =
  \mathbf{G}^{+} \mathbf{G}^{+\top}
    \bigl(
      \eta_{\mathbf{U}_i}^\top \mathbf{U}_i^{+}
      + \mathbf{U}_i^{+\top} \eta_{\mathbf{U}_i}
    \bigr)
  \mathbf{G}^{+} \mathbf{G}^{+\top}.
  \label{eq:lyapunov}
\end{equation}
Define
\begin{equation}
  \zeta_{\mathbf{U}_i}
  :=
  \eta_{\mathbf{U}_i}
  - \mathbf{U}_i^{+} \mathbf{S}_{\mathbf{U}_i}^{+}
    \bigl(\mathbf{G}_i^{+} \mathbf{G}_i^{+\top}\bigr)^{-1},
  \qquad
  \zeta_\mathbf{\underline{G}} := \eta_\mathbf{\underline{G}}.
  \label{eq:tangent-proj}
\end{equation}
Then
\[
  \zeta = (\zeta_{\mathbf{U}_1}, \dots, \zeta_{\mathbf{U}_d}, \zeta_\mathbf{\underline{G}})
  \in T_{x_{k+1}}\mathcal{M}.
\]
A cheaper variant (often used in practice) is to take $\zeta$ as the search direction,
\[
  T^{(\text{cheap})}_{x_k \to x_{k+1}}(\eta_k)
  := \zeta,
\]
i.e., only the tangent projection at the new point.

\paragraph*{\rm\textbf{Line search}}

At iteration $k$, let
\[
  x_k = (\mathbf{U}_1, \dots, \mathbf{U}_d, \mathbf{\underline{G}}), \qquad
  \operatorname{grad} L(x_k) = (g_{\mathbf{U}_1}, \dots, g_{\mathbf{U}_d}, g_{\mathbf{\underline{G}}})
\]
be the Riemannian gradient, and let
\[
  \eta_k = (\eta_{\mathbf{U}_1}, \dots, \eta_{\mathbf{U}_d}, \eta_{\mathbf{\underline{G}}}) \in T_{x_k}\mathcal{M}
\]
be the search direction (horizontal).

Define the 1D function
\begin{equation}
  \varphi_k(\alpha) := L\bigl(R_{x_k}(\alpha \eta_k)\bigr).
  \label{eq:phi-def}
\end{equation}
The directional derivative at $\alpha = 0$ is
\begin{equation}
  \varphi_k'(0)
  = g_{x_k}\bigl(\operatorname{grad} L(x_k), \eta_k\bigr)
  = \sum_{i=1}^d
    \operatorname{tr}\bigl(
      g_{\mathbf{U}_i}^\top \eta_{\mathbf{U}_i} \mathbf{G}_i \mathbf{G}_i^\top
    \bigr)
    + \langle g_{\mathbf{\underline{G}}}, \eta_{\mathbf{\underline{G}}} \rangle_F.
  \label{eq:phi-deriv}
\end{equation}

A standard Armijo backtracking line search is as follows:
\begin{itemize}
  \item Choose constants $c \in (0,1)$ (e.g.\ $c = 10^{-4}$) and
    $\tau \in (0,1)$ (e.g.\ $\tau = 1/2$).
  \item Initialize $\alpha = \alpha_0 > 0$ (e.g.\ $\alpha_0 = 1$).
  \item While
    \[
      \varphi_k(\alpha)
      >
      L(x_k) + c \,\alpha \,\varphi_k'(0),
    \]
    set $\alpha \leftarrow \tau \alpha$.
  \item Accept the final $\alpha$ as the step size $\alpha_k$.
\end{itemize}
\section{DNS Validation Cases and Synthetic Particle Track Generation}\label{App:DNS Validations}
\subsection{Forced Isotropic Turbulence}
The first validation case uses synthetic particle tracks generated from the Direct numerical simulation (DNS) velocity field of forced isotropic turbulence obtained from the Johns Hopkins Turbulence Database~\cite{minping2012forced}. 
The DNS is performed in a periodic domain of size $2\pi \times2\pi\times 2\pi$, discretized into a $1024^3$ grid. 
The Taylor micro scale Reynolds number of the flow is 433. The time separation between consecutive snapshots of the data is $\Delta t=0.002$. 
The synthetic particle tracks are generated by advecting uniformly distributed particles in the domain using a fourth-order Range-Kutta time integration scheme. 

The region of interest for  validation is a $128\times128\times50$~-voxel sub-volume where the reconstruction is performed using particle tracks of two different particle densities-- i)~0.01 particle per voxel ($\approx 8000$ particles per frame) and ii)~0.04 particle per voxel ($\approx 32000$ particles per frame). 
The two data sets represent 1\% and 4\% available data for reconstruction, respectively, compared to the number of grid points in the domain. 
In both cases, reconstruction is performed using data extracted from  $[128 \times 128 \times 50 \times 40]$ spatio-temporal section.

\begin{figure}[htbp]
\centering
\includegraphics[width=1\textwidth]{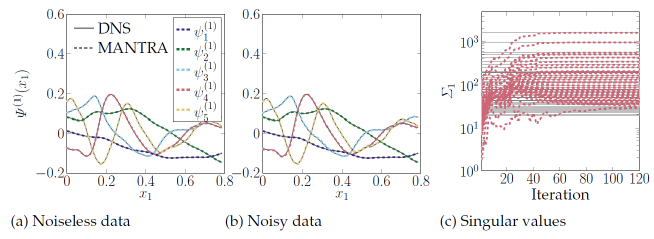}
    \caption{Isotropic turbulence: recovery of the $x_1$ correlation structure from sparse data with 4\% spatial data density. 
    (a) First five DNS correlation eigenfunctions (solid) and reconstructed eigenfunctions (dashed) without added noise. 
    (b) Corresponding eigenfunctions with 10\% added Gaussian noise. 
    (c) Convergence of the leading singular values during optimization toward the DNS values.}
    \label{fig:Adaptive_bases_and_correlation_p040}
\end{figure}

\subsection{Flow Over a Cylinder}
The second validation case is conducted using synthetic particle tracks obtained from the DNS data of flow over a cylinder. The Reynolds number of the flow is 3900. The DNS data is located in a domain of $x\times y\times z \in[4D, 8D]\times[3D, 5D] \times[2D, 4D]$, where $D$ is the diameter of the cylinder. The domain is discretized into a $309\times328\times87$ grid and the time separation between each frame is 0.00075~$D/U_{\infty}$. Synthetic particle tracks are generated by introducing 200,000 particles in the domain which are transported in time using a fourth order Range-Kutta scheme. The velocities of the synthetic particles are estimated by tri-linear interpolation from nearest eight grid points. The details about the DNS and synthetic track generation can be found in~\cite{khojasteh2022lagrangian}. 

The validation data set is formed by  collecting 21 frames in time with a time interval of 0.00375~$D/U_{\infty}$. Each frame contains approximately 200,000 particles tracks (2\% data compared to the number of DNS grid points).

\bibliography{Hessam,FTLRA}
\end{document}